\documentclass[a4paper,twoside]{article}
\usepackage{epsfig}

\newcommand{\affil}[1]{$^{\rm #1}$}

\usepackage[authoryear]{natbib}
\bibpunct{(}{)}{;}{a}{}{,}
\usepackage{graphicx}
\usepackage{epstopdf} 

\date{} 

\title{\large\bf\flushleft The Angular Diameter and Fundamental Parameters of Sirius
A}
\author{\parbox{\textwidth}{\flushleft
\vspace{-0.5cm}
{\it J.~Davis\affil{A,B}, M.~J.~Ireland\affil{A}, J.~R.~North\affil{A}, J.~G.~Robertson\affil{A,C}, W.~J.~Tango\affil{A},\\
P.~G.~Tuthill\affil{A}}\\
\vspace{0.4cm}
{\small \affil{A}\,Sydney Institute for Astronomy, School of Physics, University of
  Sydney, NSW 2006, Australia}\\
{\small \affil{B}\,Deceased 2010 January 15}\\
{\small \affil{C}\,Email: Gordon.Robertson@sydney.edu.au}}}
\begin{document}
\twocolumn[
\begin{changemargin}{.8cm}{.5cm}
\begin{minipage}{.9\textwidth}
\vspace{-1cm}
\maketitle
%
%
\small{\bf Abstract:} The Sydney University Stellar Interferometer (SUSI) has been used to make a new determination of the angular diameter of Sirius A.
The observations were made at an effective wavelength of 694.1\,nm and the
new value for the limb-darkened angular diameter is 6.048$\pm$0.040\,mas ($\pm$0.66\,\%).  This new result is compared with previous measurements and is found to be in excellent agreement with a conventionally calibrated measurement made with the European Southern Observatory's Very Large Telescope Interferometer (VLTI) at 2.176\,$\mu$m (but not with a second globally calibrated VLTI measurement). A weighted mean of the SUSI and first VLTI results gives the limb-darkened angular diameter of Sirius A as
6.041$\pm$0.017\,mas ($\pm$0.28\,\%).  Combination with the
Hipparcos parallax gives the radius equal to 1.713$\pm$0.009\,$R_{\odot}$.
The bolometric flux has been determined from published photometry and
spectrophotometry and, combined with the angular diameter, yields the emergent
flux at the stellar surface equal to (5.32$\pm$0.14)$\times10^{8}$\,Wm$^{-2}$
and the effective temperature equal to 9845$\pm$64\,K.  The luminosity is
24.7$\pm$0.7\,$L_{\odot}$.


\medskip
\medskip
\end{minipage}
\end{changemargin}
]
\small

\section{Introduction}

Sirius A (HR 2491, HD 48915) is the brightest star in the sky
($V$~=~-1.46).  It has a spectral classification of A1\,V and it is also
a member of a binary system with a less massive companion Sirius B,
first predicted by \citet{bessel}, that is now known to be a white
dwarf.  The angular diameter of Sirius A was first determined
by \citet{56sirius} in an experiment to demonstrate the potential
of intensity interferometry to overcome the problems of seeing and
extreme mechanical stability that had prevented the development of
the Michelson interferometer at that time.  \citet{58sirius}
refined their analysis to give the angular diameter of Sirius A as
7.1$\pm$0.55\,mas (pe).  While this value is now only of historical
interest, this successful experiment led to the development of the
Narrabri Stellar Intensity Interferometer (NSII) \citep*{67hbda}.
Since the pioneering measurement by Hanbury Brown \& Twiss the
angular diameter of Sirius A has been measured with long baseline
optical/infrared interferometers with increasing accuracy.  These
measurements include those by \citet*{74hbda} with the NSII, by
\citet{86dt} with the prototype for the Sydney University Stellar
Interferometer (SUSI), by \citet{mk3} with the Mark III
interferometer, and by \citet{03ketal} and \citet*{09esocals} with
the European Southern Observatory's Very Large Telescope
Interferometer (VLTI). It should be noted that in all these
measurements the white dwarf companion is too faint to have any influence
on the results for the angular size of the primary.

\citet{85hayes} reviewed the status of stellar absolute fluxes and
energy distributions from 0.32 to 4.0\,$\mu$m and included the
energy distribution of Vega (HR 7001, HD 172167) from
330--1050\,nm.  This energy distribution for Vega has been
generally accepted as the primary spectrophotometric standard for
the visual region of the spectrum but Vega's dust shell creates
problems in the infrared. \citet{92cohen} have calibrated a model
atmosphere for Vega with Hayes' energy distribution and used
infrared photometry differentially to establish an absolute scale
for a model atmosphere for Sirius~A. They derive an angular
diameter of 6.04\,mas for Sirius~A and advocate the use of
Sirius~A as the primary infrared standard beyond 20\,$\mu$m
because of Vega's dust shell.  It follows that the angular
diameter of Sirius A has assumed particular importance.

The most recent measurements of the angular diameter of Sirius A
with the VLTI \citep{03ketal, 09esocals} have by far the smallest uncertainties (by at least a
factor of 4) but the values for the angular diameter are larger
than the preceding values and the two VLTI values differ by more
than twice their standard errors. Because of the importance of the
angular diameter of Sirius A we have made an
independent measurement with SUSI using the relatively new red
beam-combination system \citep{07susi} with its improved
calibration and seeing correction techniques compared with the
initial blue-sensitive system.  In Section~\ref{sec:obs} the
observations and their calibration are described.  The detailed
analysis of the resulting calibrated values of fringe
visibility squared ($V^{2}$) is described in
Section~\ref{sec:analysis} and the results are compared with
previous measurements of the angular diameter of Sirius A in
Section~\ref{sec:discuss}.  Fundamental parameters for Sirius A
are derived and presented in Section~\ref{sec:params}.

\section{The Observations} \label{sec:obs}

The observations were made on 4 nights in March-April 2007 with
SUSI \citep{99susi1} using the red beam-combination system
\citep{07susi} operating at a nominal wavelength of 700\,nm with
an 80\,nm spectral bandwidth.  A total of 24 observations, listed
in Table~\ref{tab:obs}, were made with baselines of 5, 10, 15 and
20\,m.

\begin{table}[t]
\begin{center}
  \caption{Calibrated visibility squared ($V^{2}_{\mathrm{*cal}}$) measurements for Sirius as a function of date
  in 2007. The baseline is the projected baseline.  The values of $V^{2}_{\mathrm{*cal}}$
  are the values obtained after omitting $\eta$~CMa from the calibration process (see Section~\ref{sec:analysis}).}
  \label{tab:obs}
  \vspace{2mm}
  \begin{tabular}{lcccl}
  \hline
 \multicolumn{1}{c}{Date} & Zenith & Baseline & $V^{2}_{\mathrm{*cal}}$  \\
    & Angle &  &   \\
    & (deg) & (m) &  \\
 \hline
14 March & 20.1 & 14.64 & 0.388$\pm$0.018 \\
& 22.4 & 14.67 & 0.388$\pm$0.018 \\
& 24.7 & 14.70 & 0.394$\pm$0.018 \\
& 27.1 & 14.73 & 0.391$\pm$0.021 \\
& 29.9 & 14.77 & 0.406$\pm$0.023 \\
& 32.7 & 14.81 & 0.375$\pm$0.030 \\
15 March & 13.6 & 19.44 & 0.156$\pm$0.009 \\
& 13.8 & 19.44 & 0.152$\pm$0.007 \\
& 14.5 & 19.45 & 0.163$\pm$0.007 \\
& 24.2 & 19.59 & 0.146$\pm$0.018 \\
& 26.9 & 19.64 & 0.151$\pm$0.016 \\
& 29.5 & 19.69 & 0.153$\pm$0.018 \\
12 April & 20.3 & 4.88 & 0.938$\pm$0.025 \\
& 22.8 & 4.89 & 0.889$\pm$0.026 \\
& 25.0 & 4.90 & 0.917$\pm$0.025 \\
& 27.4 & 4.91 & 0.909$\pm$0.022 \\
& 30.3 & 4.93 & 0.884$\pm$0.022 \\
& 35.0 & 4.95 & 0.929$\pm$0.028 \\
& 38.4 & 4.96 & 0.936$\pm$0.024 \\
& 41.3 & 4.97 & 0.955$\pm$0.030 \\
15 April & 36.1 & 9.90 & 0.669$\pm$0.018 \\
& 39.0 & 9.93 & 0.656$\pm$0.025 \\
& 41.7 & 9.95 & 0.678$\pm$0.021 \\
& 47.3 & 9.96 & 0.630$\pm$0.021 \\
 \hline
 \end{tabular}
\end{center}
\end{table}

\subsection{Calibration} \label{sec:cal}

Calibration of the observed values of $V^{2}$ is essential to
minimise the effects of instrumental visibility losses and residual seeing
effects.  Each observation of Sirius was bracketed by observations
of two of the three selected calibrators that are listed in Table~\ref{tab:cals}.
After the initial analysis involving all three calibrators, which is
described here, it was found that $\eta$~CMa was not a satisfactory
calibrator.  The discovery and subsequent calibration omitting $\eta$~CMa
is described in Section~\ref{sec:analysis}.
The angular diameters of all three calibrators were measured with the NSII.
They are all small compared with the angular diameter of Sirius A so that
the uncertainties in their expected $V^{2}$ values, at the baselines
employed, should be essentially negligible compared with the
uncertainties in the measured values of $V^{2}$.
The effective wavelength of observations for all
three calibrators is estimated to be 693$\pm$2\,nm following the
procedure described by \citet{07susi}.  The uniform-disk angular diameters at
693\,nm have been derived from the limb-darkened angular diameters
given by \citet{74hbda} using the appropriate correction factors
from the tabulation by \citet{00dtb} which were derived from
Kurucz model atmosphere centre-to-limb intensity distributions
(\citealt{93akurucz}, \citealt{93bkurucz}).  The uncertainties in the
uniform-disk angular diameters of the calibrators have been taken into
account in the analysis of the observational data.

\begin{table*}
\begin{center}
  \caption{The three calibrators used for the observations of Sirius.  All three stars had
  their angular diameters measured with the Narrabri Stellar Intensity Interferometer
  \citep{74hbda}.  The published limb-darkened angular diameters
  have been converted to uniform-disk angular diameters using
  correction factors interpolated from the tabulation of
  \citet{00dtb}.  The angle $\alpha$ is the great circle distance on the sky of the
  calibrator from Sirius.}
  \label{tab:cals}
  \vspace{2mm}
  \begin{tabular}{clcccr}
  \hline
 HR & \multicolumn{1}{c}{Name} & Spectral & $V$ & $\theta_{\mathrm{UD}}$ & \multicolumn{1}{c}{$\alpha$} \\
    &  & Type & & (mas) & \multicolumn{1}{c}{(deg)} \\
 \hline
 2004 & $\kappa$~Ori & B0.5\,Ia & 2.06 & 0.44$\pm$0.03 & 15.5 \\
 2294 & $\beta$~CMa & B1\,II-III & 1.97 & 0.51$\pm$0.03 & 5.4 \\
 2827 & $\eta$~CMa  & B5\,I & 2.46 & 0.73$\pm$0.06 & 15.3 \\
 \hline
 \end{tabular}
\end{center}
\end{table*}

The calibration is achieved using a transfer function ($T$) for
each observation of a calibrator defined by

\begin{equation}
T = \frac{V^{2}_{\mathrm{obs}}}{V^{2}_{\mathrm{exp}}}
\end{equation}
where $V^{2}_{\mathrm{obs}}$ is the observed value of $V^{2}$ and
$V^{2}_{\mathrm{exp}}$ is the expected value of $V^{2}$ computed
from the projected baseline of the observation and the
uniform-disk angular diameter.

Details of how the $T$ values have been used are given in
Section~\ref{sec:analysis}.

\subsection{Observational Technique} \label{sec:technique}

Details of the standard SUSI observing
technique have been described by \citet{07susi}.  In brief, for
the observations of Sirius A and its calibrators, each observation
consisted of a set of 1000 scans through the fringe envelope,
together with photometric measurements for normalisation purposes.
Each observation took approximately 6\,minutes so each bracketed
observation of Sirius A was completed in $\sim$18\,minutes.
Although it was impossible to match the range in hour angle (and hence
zenith angle) exactly for the four baselines, they were matched as
far as possible and the mean zenith angle of each observation is
listed in Table~\ref{tab:obs}.  Automatic alignment checks were
carried out at intervals of approximately one hour as described
by \citet{07susi}.

\section{The Analysis}\label{sec:analysis}

The observational data were initially processed with the SUSI
pipeline as described by \citet{07susi} which outputs a value of
$V^{2}$ for each observation, appropriately flux normalised and
seeing corrected.  Subsequent analysis was carried out using an MS
Excel spreadsheet.  The first step was to compute $T$ for each
calibrator observation.  Next the weighted mean
$\overline{T}$ for each pair of bracketing calibrators
was found and the intervening value of $V^{2}$ for Sirius A was
divided by the mean $T$ to obtain a calibrated value of $V^{2}$ as
given by

\begin{equation}
V^{2}_{\mathrm{*cal}} =
\frac{V^{2}_{\mathrm{*obs}}}{\overline{T}}
\end{equation}
where $V^{2}_{\mathrm{*obs}}$ is the value of $V^{2}$ for Sirius A
from the pipeline and $V^{2}_{\mathrm{*cal}}$ is the calibrated
value of $V^{2}$ for Sirius A.

An initial fit to the data to determine the uniform-disk angular
diameter as described in Section~\ref{sec:udad} gave a zero baseline
value for $V^{2}_{\mathrm{*cal}}$ of $A = 1.0348\pm0.0094$, 3.7$\sigma$
greater than the expected value of unity.  An investigation was therefore
made of the calibrators by inter-comparing their transfer functions.
For each pair of calibrators bracketing a Sirius observation their
transfer functions ($T$) were plotted against each other.  In the case
of the plot for $\beta$~CMa versus $\kappa$~Ori the slope is
1.000$\pm$0.013 showing that these two calibrators are in good agreement,
with the scatter about the fitted line consistent at all four baselines.
The corresponding plots of $T$ for $\eta$~CMa against $\beta$~CMa and against
$\kappa$~Ori on the other hand show a different slope for each baseline,
in all cases significantly different from unity.  This suggests a previously
unsuspected binary nature for $\eta$~CMa.  Without a detailed orbit for
$\eta$~CMa it is not possible to use this star as a calibrator.  Observations
of $\eta$~CMa have therefore been ignored and the following procedure adopted.
Where $\eta$~CMa was observed, the weighted mean value of $T$ for the
$\beta$~CMa and $\kappa$~Ori on either side of it was adopted.  The two intervening
values of $V^2_{\mathrm{*cal}}$ have the same calibration value of $T$.  This
approach was possible for all except two Sirius observations at 20\,m and these have
been omitted.  The intervals between the remaining calibrator observations are increased
from $\sim$12\,minutes to $\sim$24\,minutes but this is acceptable.

The calibrated values of $V^{2}$ for Sirius A (i.e.
$V^{2}_{\mathrm{*cal}}$) are listed in Table~\ref{tab:obs}.

\subsection{The Uniform-Disk Angular Diameter} \label{sec:udad}

The uniform-disk angular diameter is obtained by means of a least
squares fit of the following expression to the calibrated values
of $V^{2}$ listed in Table~\ref{tab:obs}

\begin{equation}
V^{2}_{\mathrm{*cal}} = A\left|\frac{2J_{1}(x)}{x}\right|^{2}
\label{eqn:bessel}
\end{equation}
where $J_{1}(x)$ is a Bessel function, $x = \pi
b\theta_{\mathrm{UD}}/\lambda_{\mathrm{eff}}$, with $b$ the
projected baseline, $\theta_{\mathrm{UD}}$ the uniform-disk
angular diameter, and $\lambda_{\mathrm{eff}}$ the effective
wavelength of observation. $A$ is the value of $V^{2}$ obtained
by extrapolating the fitted curve to zero baseline.  The fit is
therefore a two parameter fit for the uniform-disk angular
diameter ($\theta_{\mathrm{UD}}$) and the zero baseline value of
$V^{2}_{\mathrm{*cal}}$ ($A$).

The effective wavelength for observations of main-sequence stars
using SUSI's 700\,nm filters has been studied by \citet{07susi}, where
two methods were used: (i) calculation based on the known filter bandpass, stellar spectral distributions and instrumental sensitivity as a function of wavelength, and (ii) measurements based on the fringe frequency derived from laser-calibrated scans. Rather than simply take the value from 
the tabulation of \citet{07susi}, we here make use of the fact that Sirius was one of the stars for which a measurement was made.
%
%
The measured value was 693.69$\pm$0.34\,nm while the calculated value was 694.5\,nm. Owing to the scatter in the measured values of $\lambda_{\mathrm{eff}}$, we err on the side of caution, and adopt a
value of 694.1\,nm, which is midway between the computed and measured
values, but with an uncertainty of $\pm1.0$\,nm to allow for the
uncertainties in the measured and computed values for this star.

The relatively broad spectral bandwidth may lead to bandwidth smearing.  The resolution of the interferometer will
vary across the band and limb-darkening, which is wavelength
dependent, will also vary across the band.  These effects have
been discussed by \citet{02td} and have been investigated
specifically for the current observations of Sirius A.  Bandwidth
smearing is not significant in this case and the conclusion is
that equation~(\ref{eqn:bessel}) can be fitted to the
observational data using the effective wavelength as discussed
above.

A fit of equation~(\ref{eqn:bessel}) to the individual values of
$V^{2}_{\mathrm{*cal}}$ in Table~\ref{tab:obs} gives $A =
1.0151\pm0.0093$ ($\pm$0.9\,\%) and $\theta_{\mathrm{UD694.1}} =
5.872\pm0.034\,\mathrm{mas}$ ($\pm$0.6\,\%) with a reduced $\chi^{2}
= 0.63$.  Although the fit gives a zero baseline value of $A$ that is ~1.6\,$\sigma$
greater than the expected value of unity this is regarded as satisfactory
in view of the comparison of the two calibrators that were adopted.

\begin{figure}[t]
\begin{center}
\includegraphics[scale=0.55]{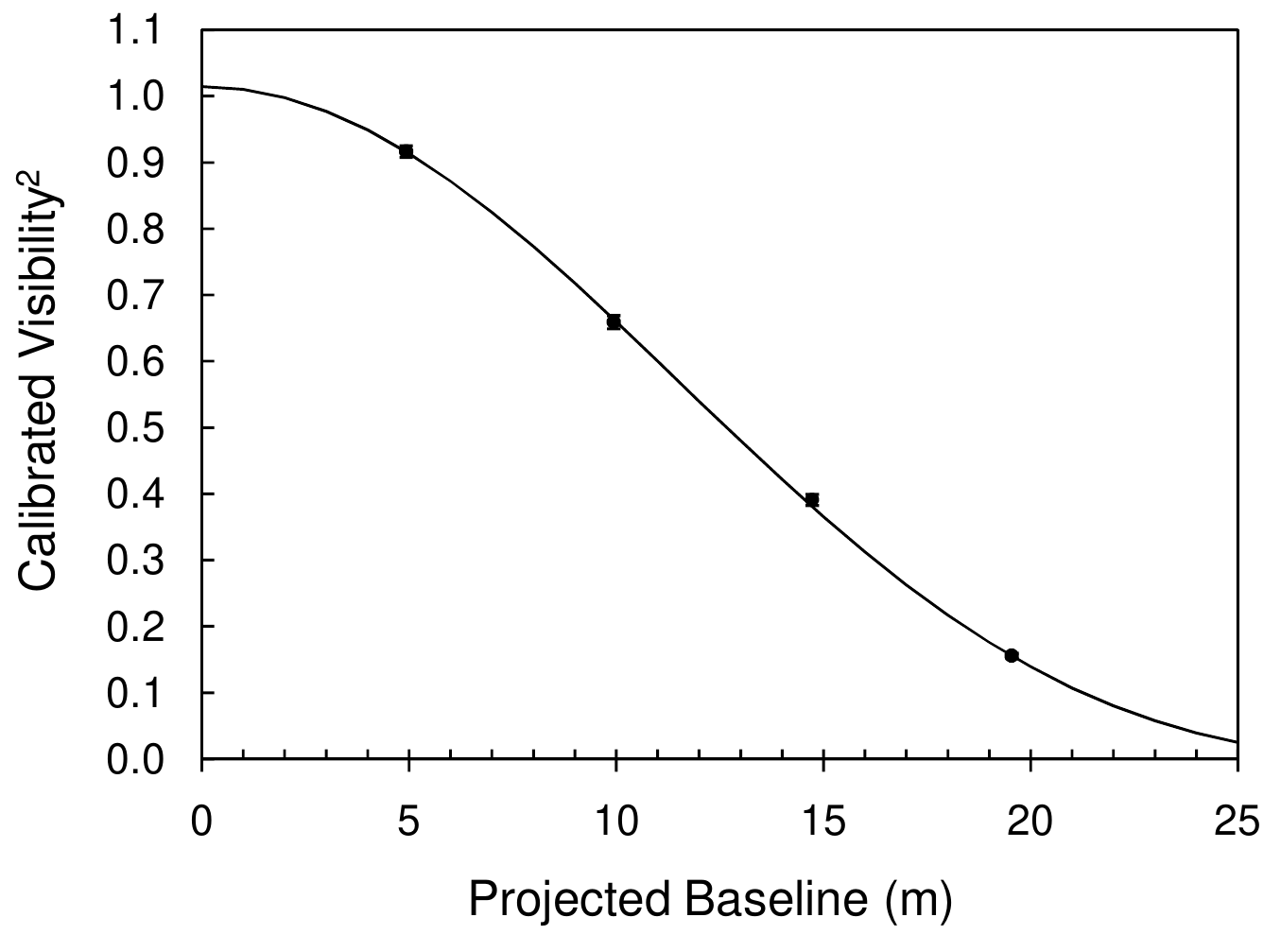}
  \caption{The weighted mean values of calibrated squared visibility at the
  four baselines used for the observations.  The line is the fit to the individual values
  of observational data listed in Table~\ref{tab:obs}.}
  \label{fig:transform}
\end{center}
\end{figure}

The uniform-disk angular diameter, taking the uncertainty in the effective wavelength
into account, is 5.872$\pm$0.038\,mas.  The fit to all the observational data is shown
in Figure~\ref{fig:transform}.  As can be seen the fit to all four baselines is excellent.

\begin{figure}[t]
\begin{center}
\includegraphics[scale=0.5]{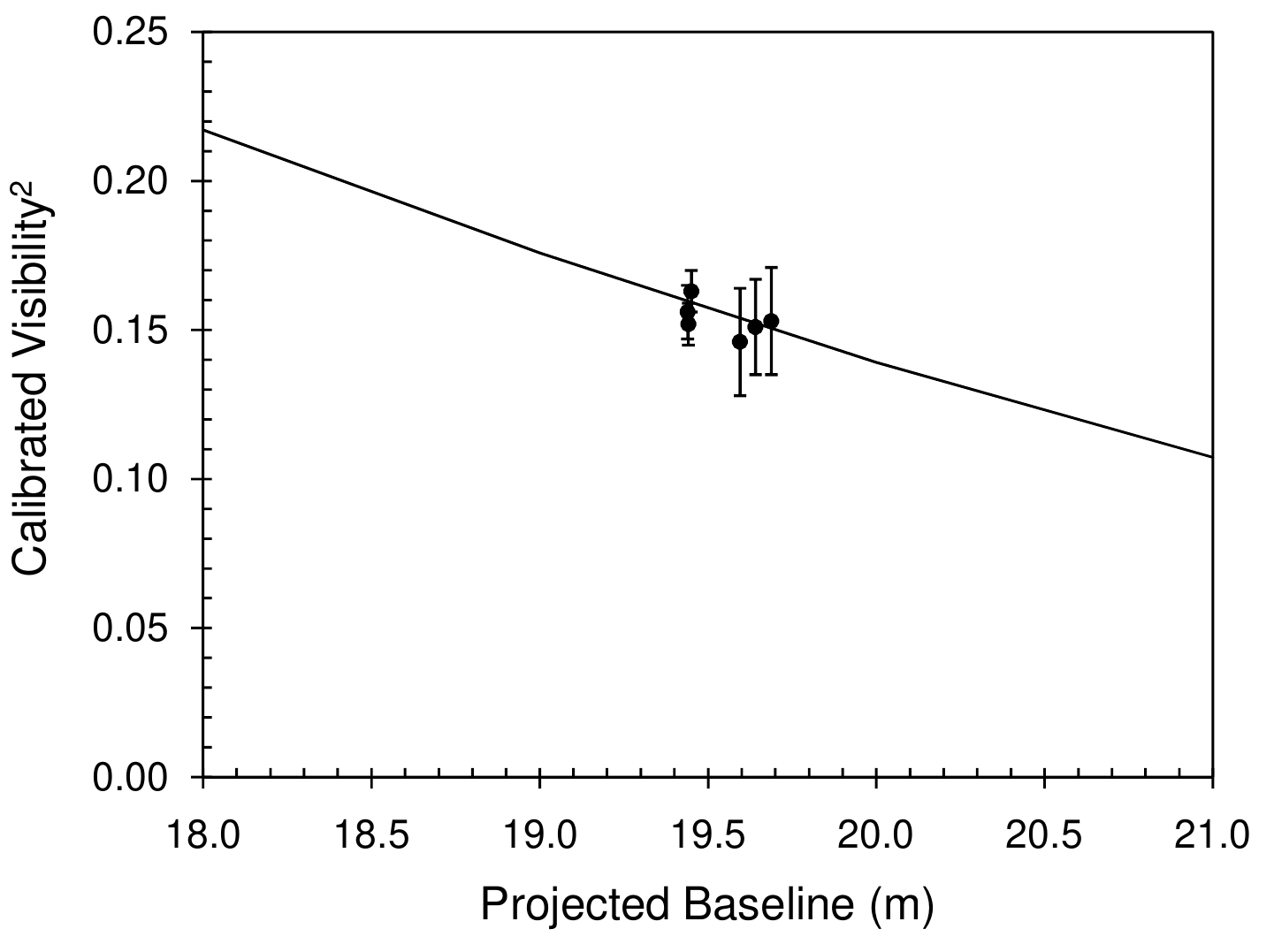}
  \caption{The individual calibrated squared visibilities at the longest baseline.
  Although all six measurements were made on the same night, the seeing deteriorated
  significantly between the first three and last three measurements as illustrated
  by the larger error bars.}
  \label{fig:long}
\end{center}
\end{figure}

The 8 individual values of $V^{2}_{\mathrm{*cal}}$ for the 20\,m
baseline are shown in Figure~\ref{fig:long} and show the small
scatter in the data.

\subsection{Limb-darkening Correction}

In order to obtain the true, limb-darkened angular diameter a
wavelength-dependent correction has to be applied to the
uniform-disk angular diameter.  \citet{00dtb} have tabulated
correction factors, derived from Kurucz model atmosphere
centre-to-limb intensity distributions (\citet{93akurucz},
\citet{93bkurucz}), as a function of wavelength, effective
temperature, surface gravity, and [Fe/H].

For Sirius A the following parameters have been adopted following
\citet{92cohen}: $T_{\mathrm{eff}} = 9850$\,K, $\log{g} = 4.25$,
and [Fe/H] = $+0.5$.  With the effective wavelength of 694.1\,nm,
interpolation in the \citet{00dtb} tabulation gives
$\rho_{694.1}$ = 1.030.  Although the interpolation itself is
accurate to 0.0001 we adopt an uncertainty of 0.001 because the Kurucz
models used by \citet{00dtb} may not be precisely representative of Sirius A.

The limb-darkened angular diameter is therefore
6.048$\pm$0.040\,mas.

\section{The Limb-darkened Angular Diameter of Sirius A} \label{sec:discuss}

\begin{figure}[t]
\begin{center}
\includegraphics[scale=0.56]{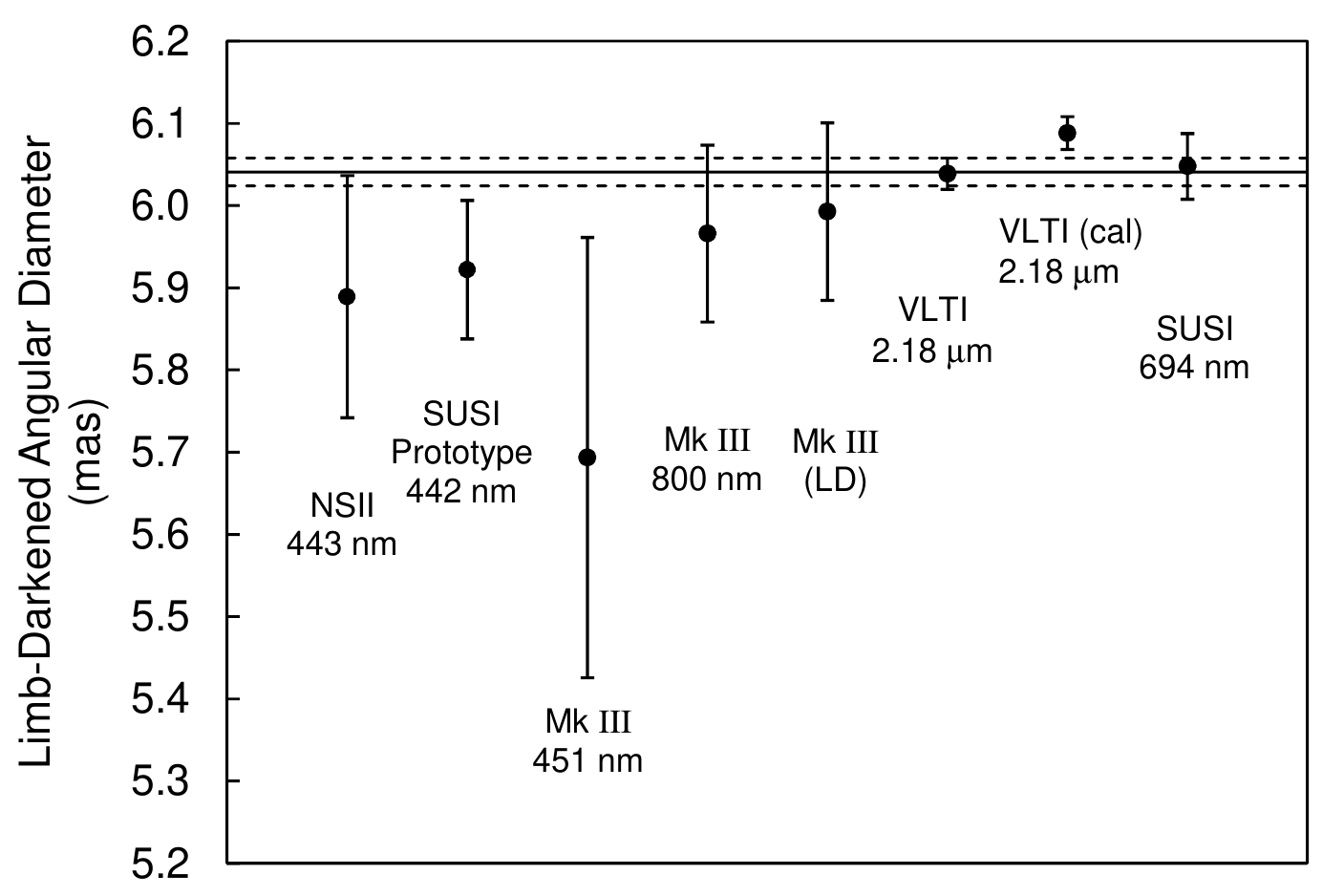}
  \caption{The limb-darkened angular diameter determinations for
Sirius listed in Table~\ref{tab:ldads}, except for the original
historical measurement of \citet{58sirius}, in order of
publication. The horizontal line is the weighted mean of the VLTI and SUSI
values and the dashed lines represent the uncertainty in the
mean.  The VLTI(cal) value is from Richichi et al. (2009).}
  \label{fig:angdiams}
\end{center}
\end{figure}

Table~\ref{tab:ldads} lists the measurements of the angular
diameter of Sirius A made to date and Figure~\ref{fig:angdiams}
plots the tabulated values of limb-darkened angular diameter in
order of measurement.

\begin{table*}
\begin{center}
  \caption{Measurements of the limb-darkened angular diameter of Sirius.  Entries marked with an asterisk are
  published uniform-disk angular diameters that have been converted to limb-darkened angular diameters
  with the limb-darkening factor $\rho_{\lambda}$ interpolated from the tabulation by \citet{00dtb}.  In
  the case of the published Mark III values a separate limb-darkened angular diameter was also included \citep{mk3}.}
  \label{tab:ldads}
  \begin{tabular}{clccccc}
  \hline
 Year & \multicolumn{1}{c}{Instrument} & $\lambda$ & $\theta_{\mathrm{UD}}$ & $\rho_{\lambda}$ & $\theta_{\mathrm{LD}}$ & \multicolumn{1}{c}{Ref.} \\
    &  & (nm)& (mas)  &  & (mas)     & \\
 \hline
 1958 & NSII Prototype   & 500   & -- & -- & 7.1$\pm$0.9      & 1  \\
 1974 & NSII             & 443.0 & 5.60$\pm$0.15& 1.052 & 5.89$\pm$0.15    & 2  \\
 1986 & SUSI Prototype * & 441.6 & 5.63$\pm$0.08 & 1.052 & 5.92$\pm$0.08    & 3  \\
 2003 & Mark III *       & 800 & 5.823$\pm$0.105 & 1.025 & 5.966$\pm$0.108  & 4  \\
 2003 & Mark III *       & 451 & 5.421$\pm$0.255 & 1.050 & 5.694$\pm$0.268  & 4  \\
 2003 & Mark III         &  800     & -- & -- & 5.993$\pm$0.108  & 4 \\
 2003 & VLTI (single cal)       &  K band & -- & -- & 6.039$\pm$0.019  & 5  \\
 2009 & VLTI (global cal)       & 2181  & 6.030$\pm$0.020 & 1.0097 & 6.089$\pm$0.020 & 6 \\
 2010 & SUSI             & 694.1 & 5.872$\pm$0.038 & 1.030 & 6.048$\pm$0.040  & 7  \\
 \hline
 \end{tabular}
 \end{center}
\medskip
References: (1) \citet{58sirius}; (2) \citet{74hbda}:  (3) \citet{86dt}; (4) \citet{mk3}; (5) \citet{03ketal}; (6) \citet{09esocals}; (7) This work.
\end{table*}

Sirius was a difficult target for the Mark III interferometer because it
has a minimum zenith angle of $\sim$51$^{\circ}$ at Mount
Wilson.  Seeing introduces increasing uncertainties in the
measurement of $V^{2}$ with zenith angle for amplitude
interferometry, particularly in the blue, and this is shown by the
larger uncertainty for the 451\,nm measurement compared with that
at 800\,nm.

The VLTI measurement of Sirius by \citet{03ketal} achieved an accuracy of
$\pm$0.31\,\% in spite of using $\theta$~Cen as the sole calibrator
for the long baseline ($\sim$62\,m) observations.  $\theta$~Cen is not an ideal
calibrator because it has a relatively large angular diameter
and is $\sim$96$^{\circ}$ distant from Sirius.  However, these disadvantages
appear to have been overcome by the fact that the VLTI/VINCI instrumental
combination is significantly less prone to seeing effects as a result
of the spatial filtering by the optical fiber coupling in VINCI.  As will be
seen from Table~\ref{tab:ldads} this result and the new SUSI measurement are
in excellent agreement.

The VLTI determination of the angular diameter of Sirius by \citet{09esocals}
differs from the \citet{03ketal} and new SUSI measurements in that it was not
a direct measurement based on individually calibrated values of $V^{2}$.  It was
the result of a global fit to some 700 values of $V^{2}$ determined over 50 nights for
17 stars selected as `primary' calibrators.  The
analysis was based on the assumption that the transfer function remained constant
during each night and criteria were applied to select good quality data.  In the
case of Sirius, 26 separate observations spread over 6 nights were involved in
obtaining the result listed in Table~\ref{tab:ldads}.  The resulting limb-darkened
angular diameter is larger than that of \citet{03ketal} as noted by \citet{09esocals}
and also larger than the SUSI value.  It lies more than twice its standard error
from both.  Although the formal uncertainty is small there are concerns about the
convergence of the algorithm used by \citet{09esocals} and the validity of the
assumption of a constant transfer function for each night of observation.  These
limitations were recognised and discussed by Richichi et al. and, while in general
their results were in good agreement with the results of other
measurements for stars in common, they noted that Sirius was an exception.

It is desirable to combine measurements to determine a weighted mean value for
the limb-darkened angular diameter of Sirius but this poses the question of
which values to accept.  Examination of the values listed in Table~\ref{tab:ldads}
shows that the early values, including the Mark III result, have relatively low
weights and that they would have little influence on a weighted mean value.
The remaining values are the two VLTI results and the new SUSI value.  The
concerns expressed regarding the \citet{09esocals} VLTI result, and the fact
that it lies significantly above the other two values, suggests that the formal
uncertainty in this case is underestimated and, furthermore, that some undetermined systematic effect may exist.  One possibility would be to
arbitrarily increase the uncertainty but, since the other two values are in
good agreement, a weighted mean of the \citet{03ketal} VLTI and SUSI results has been
adopted.  The weighted mean of these two results for the limb-darkened angular
diameter of Sirius A is 6.041$\pm$0.017\,mas.  The value of 6.04\,mas determined
by \citet{92cohen} by spectral radiance calibration is in excellent agreement with
the value determined by direct measurement.

\section{The Parameters of Sirius A} \label{sec:params}

The combination of the limb-darkened angular diameter with the
bolometric flux received from a star and with its parallax enables
a number of important stellar parameters to be determined.  First
the bolometric flux received from the star must be established.

\subsection{The Bolometric Flux} \label{sec:bolflux}

The integrated flux for Sirius was determined by \citet{76code} to be
(114.3$\pm$4.4)$\times10^{-9}$\,Wm$^{-2}$.  However, additional data have
become available and improvements in calibration have been achieved since
the Code et al. value was published.  Following a similar procedure to that
employed by Code et al. a new value for the bolometric flux received from Sirius
has been established.  The spectrum has been divided into four wavelength ranges
for this purpose: 0-0.33\,$\mu$m, 0.33-0.80\,$\mu$m, 0.80-5.00\,$\mu$m
and 5.00-$\infty$\,$\mu$m.  The flux beyond 5.0\,$\mu$m is $<0.1$\,per cent of the total
flux for a model stellar atmosphere for Sirius and, as discussed below, allowance is made for it.

It is noted that the ($B$-$V$) and ($U$-$B$) colours for Sirus are consistent with those of
an unreddened star of its spectral classification and no correction for interstellar extinction
has been necessary.

\subsubsection{Ultraviolet Flux}

Ultraviolet flux measurements include four photometric measurements made with the
TD-1 satellite \citep{78td1}, five photometric measurements made with the OAO-2
satellite \citep{80oao2}, and an OAO-2 spectrum \citep{79oao2}.  The latter represents by far
the most comprehensive flux data for the wavelength range 120-360\,nm, and it is in excellent
agreement with the adopted visual fluxes, discussed in the next section, in the overlap
region 330-360\,nm.  In view of this the OAO-2 spectrum has been adopted for the wavelength range
120-330\,nm and it has been integrated in three sections, 120-130\,nm, 130-180\,nm and 180-330\,nm
corresponding to the ranges given by \citet{76bless} with differing calibration uncertainties.
The uncertainties for the three ranges have been estimated following the procedure adopted by
\citet{76code} but, whereas Code et al. made estimates for the hottest stars and applied it to all stars,
here the estimates have been made specifically for Sirius A based on the observed flux distribution.
In order to estimate the uncertainty in the integrated flux for the band 130-180\,nm, the
uncertainty in the absolute flux at 130\,nm has been assumed to be $\pm$30\,\% and,
at 180\,nm, $\pm$10\,\%, varying linearly between these limits.  For the band 180-330\,nm
the uncertainty at 180\,nm has been assumed to be $\pm$10\,\% and, at 330\,nm, $\pm$5\,\%,
varying linearly in between.  For 120-130\,nm the fluxes are uncertain and, following \citet{76code}, an
uncertainty of $\pm$35\,\% has been adopted.  The integrated fluxes for the ultraviolet are:
for 120-130\,nm (0.3$\pm$0.1), for 130-180\,nm (11.0$\pm$1.9), and for 180-330\,nm (23.4$\pm$1.7), all
$\times10^{-9}$ Wm$^{-2}$.

\subsubsection{Visual Flux}

Spectrophotometric scans of the visual region of Sirius's spectrum include those by
\citet{66afn} (339.2-586.8\,nm), \citet{68gm} (339-584\,nm), \citet{71spo} (330-605\,nm),
and \citet{74dw} (330-808\,nm).  The scan by \citet{74dw} has been adopted since it covers
the greatest wavelength range.  Support for this decision is given by the excellent
agreement found for the \citet{74dw} spectrophotometry, in the case of $\delta$\,CMa \citep{07dcma},
with that of \citet{87kiehling} and the flux distribution in the MILES library of
empirical spectra \citep{06miles}.

\citet{76code} used the spectrophotometric calibration of $\alpha$~Lyr (Vega) by \citet{70oke}
to convert the relative spectrophotometry of \citet{74dw} into a relative absolute flux
distribution.  Here the more recent spectrophotometric calibration of Vega by \citet{85hayes}
has been used.  Following \citet{76code} the resulting relative absolute flux distribution
has been scaled by the flux ratio corresponding to the monochromatic magnitude of $\alpha$~CMa
relative to Vega at 550\,nm (-1.456) measured by Davis (private communication). It has
then been converted to fluxes using the value for the flux from Vega at 550.0\,nm of
$3.56\times10^{-11}$\,Wm$^{-2}$nm$^{-1}$ \citep{95meg}.

The spectrophotometric fluxes are continuum fluxes and do not allow for the strong Balmer absorption
lines.  In order to make allowance for the absorption lines and the lack of measurements between
370\,nm and 403\,nm use has been made of a model atmosphere generated specifically for Sirius by
R.~Kurucz.  The model temperature of 9850\,K is essentially the same as that derived in this
work (see Section~\ref{sec:fandt}) and is therefore appropriate for this purpose.  The model flux
was integrated for the interval 330-800\,nm and also for the model with the lines replaced by
continuum interpolated from each side of the lines.  The ratio of flux with the lines to the flux
without the lines was found to be 0.9534.  The empirical flux distribution was also integrated with
the addition of three points in the 370-403\,nm interval derived from the smoothed model flux
distribution to give an integrated flux for 330-800\,nm equal to 67.26$\times10^{-9}$\,Wm$^{-2}$ which,
after correction for the lines, becomes 64.13$\times10^{-9}$\,Wm$^{-2}$.

The uncertainty in the visual flux, based on the analysis carried out for $\delta$\,CMa \citep{07dcma},
suggests that the uncertainty in the relative absolute flux distribution
is at the 1\,\% level, the uncertainty in the monochromatic difference used for scaling the
Sirius flux distribution is estimated to be $\sim$1\,\%, and \citet{95meg} claims $\pm$0.7\,\%
for the flux calibration at 550\,nm.  These uncertainties are independent and combining
them quadratically gives a resultant uncertainty of $\pm$1.6\,\%.  In addition there is an
uncertainty associated with the use of the theoretical model atmosphere and we adopt a
final uncertainty of $\pm$2.0\,\%.  This is an improvement on the uncertainty of $\pm$4\,\%
adopted by \citet{76code} but is based on evidence that the spectrophotometric scan of
\citet{74dw} is better than thought by Code et al., the improved Vega calibration by
\citet{85hayes}, and the improved calibration by \citet{95meg}.  The flux for the wavelength
range 330-800\,nm is (64.1$\pm$1.3)$\times10^{-9}$\,Wm$^{-2}$.

\subsubsection{Infrared Flux}

The fluxes for the wavelength range 0.80-5.00\,$\mu$m have been obtained from the $I$, $J$, $H$, $K$,
$L$, and $M$ broad-band photometry of \citet{66hljetal} ($I$, $J$, $K$, $L$), \citet{80cousins}
($I$), \citet{81engels} ($J$, $H$, $K$, $L$, $M$) and \citet{74glass} ($J$, $H$, $K$, $L$).
Flux calibrations by \citet{66hlj}, \citet{98bessell}, \citet{95meg}, and \citet{81wam} have been used to
convert the photometric magnitudes to fluxes.  Multiple measurements for the same band have been
averaged and the adopted fluxes are given in Table~\ref{tab:phot}.

\begin{table}
\begin{center}
    \caption{Adopted calibrated photometric IR flux values for Sirius from the literature. The units
    for wavelength ($\lambda$) are $\mu$m and for the flux 10$^{-9}$ Wm$^{-2}\mu$m$^{-1}$.}
    \label{tab:phot}
    \vspace{2mm}
    \begin{tabular}{ccccc}
    \hline
    Band  & $\lambda$  & Flux & Source & Calibration \\
    \hline
    $I$ & 0.798 & 40.9  & 1 & a \\
    $I$ & 0.90  & 31.0  & 2 & b \\
    $J$ & 1.25  & 10.8  & 2, 3, 4 & c, a, a \\
    $H$ & 1.65  & 3.90  & 3, 4 & a \\
    $K$ & 2.2   & 1.38  & 2, 3, 4 & c, a, a \\
    $L$ & 3.4   & 0.27  & 2 & b \\
    $L$ & 3.7   & 0.205 & 3, 4 & c \\
    $M$ & 4.8   & 0.076  & 3 & d \\
    \hline
    \end{tabular}
\end{center}
\medskip
Source references: (1) \citet{80cousins}; (2) \citet{66hljetal};
(3) \citet{81engels}; (4) \citet{74glass}.\\
Calibration references: (a) \citet{98bessell}; (b) \citet{66hlj};
(c) \citet{95meg}; (d) \citet{81wam}.
\end{table}

The data in Table~\ref{tab:phot} were plotted against wavelength and a smooth curve drawn
through them.  The extended curve at the short wavelength end joined smoothly to
the visual fluxes in the 0.7-0.81\,$\mu$m range and was also in agreement with the 13-colour fluxes
of \citet{13color} in the range 0.724-1.108\,$\mu$m. The area under the curve from 0.80\,$\mu$m
to 5.0\,$\mu$m was integrated to give the continuum flux for this interval equal to
15.51$\times10^{-9}$ Wm$^{-2}$.  As for the visual region the Kurucz model was integrated with the
absorption lines and with the lines replaced by continuum interpolated from each side of the lines.
The ratio of flux with the lines to the flux without the lines was found to be 0.9797.  The integrated
empirical flux, after correction for the lines, becomes 15.20$\times10^{-9}$\,Wm$^{-2}$.

The uncertainties in the data in Table~\ref{tab:phot} have been computed from the published
uncertainties in the photometry and in the calibrations.  The average uncertainty across the
2.0-5.0\,$\mu$m band was found to be $\pm$4.1\% and this has been adopted as the uncertainty in
the integrated flux.  The integrated flux is therefore (15.2$\pm$0.6)$\times10^{-9}$\,Wm$^{-2}$.

\subsubsection{The Residual Flux for $\lambda > 5.0\,\mu$m}

As mentioned in Section~\ref{sec:bolflux} the flux from Sirius for $\lambda > 5.0\,\mu$m is
small.  The flux longward of 5\,$\mu$m to 297\,$\mu$m for the Kurucz model atmosphere for Sirius
has been integrated and is 0.09\,\% of the total flux for the model.  In the case of a black body
similar integrations give ~0.11\,\% of the total flux longward of 5.0\,$\mu$m.  Since this is such
a small fraction of the total flux the IRAS point source fluxes ($\geq$12\,$\mu$m) and the low-resolution
scan data have not been considered.  The flux longward of 5.0\,$\mu$m has been
taken to be 0.10$\pm$0.01\,\% of the total flux or (0.11$\pm$0.01)$\times10^{-9}$\,Wm$^{-2}$.

\subsubsection{The Total Flux}

The fluxes for the four wavelength ranges considered are listed in Table~\ref{tab:fluxes}
together with the bolometric flux ($f$), the total flux received from Sirius, which is equal to
(114.1$\pm$2.9)$\times10^{-9}$\,Wm$^{-2}$.  The bolometric flux is close to the value of
(114.3$\pm$4.4)$\times10^{-9}$\,Wm$^{-2}$ found by \citet{76code} but is more accurate
due to improvements in the flux calibrations.  The flux values used in the UV, visual and
IR are plotted in Figure~\ref{fig:flux} together with the flux distribution for the Kurucz
model atmosphere ($T_{\mathrm{eff}}=9850$\,K, $\log{g}=4.25$ and {Fe/H]=+0.5).  It is
stressed that the integrated empirical fluxes are essentially independent of the model
except for corrections for absorption lines not included in the photometric and spectrophotometric
spectral bands which amount to a total of 2.9\,\% of the total flux.  The model flux distribution is
that fitted to Sirius fluxes, derived from differential Sirius/Vega photometry, by \citet{92cohen}.
The model fluxes are independent of the fluxes determined here and have been included
in the figure to illustrate the excellent agreement between the current work and that of Cohen et al.

\begin{table}
\begin{center}
    \caption{Integrated fluxes for Sirius in each spectral band and the resulting bolometric flux.  The
    units for the wavelength bands ($\lambda\lambda$) are $\mu$m and for the flux and its uncertainty ($\sigma$)
    they are 10$^{-9}$\,Wm$^{-2}$.}
    \label{tab:fluxes}
    \vspace{2mm}
    \begin{tabular}{cccc}
    \hline
    Range & $\lambda\lambda$  & Flux & $\pm\sigma$ \\
    \hline
    UV & 0.12-0.13  & 0.3 & 0.1 \\
    UV & 0.13-0.18 & 11.0 & 1.9 \\
    UV & 0.18-0.33 & 23.4 & 1.7 \\
    Visual & 0.33-0.80 & 64.1 & 1.3 \\
    IR & 0.80-5.00 & 15.2 & 0.6 \\
    Far IR & $>$ 5.00  &  0.1 &  \\
    \hline
    \multicolumn{2}{l}{Bolometric Flux ($f$)} & 114.1 & 2.9 \\
    \hline
    \end{tabular}
\end{center}
\end{table}

\begin{figure*}
\begin{center}
\includegraphics[scale=0.9]{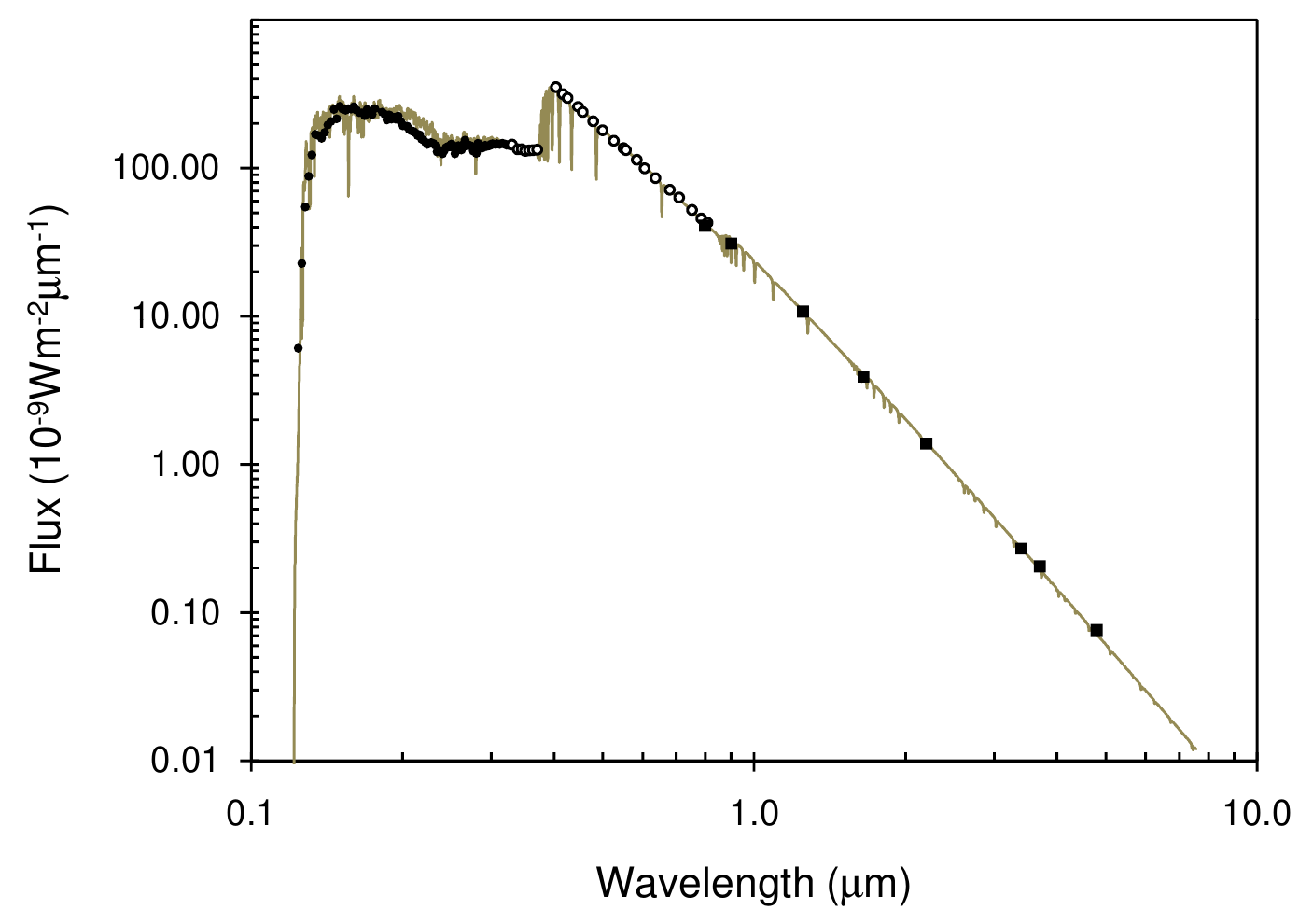}
  \caption{The flux distribution for Sirius from the ultraviolet to the mid-infrared.
  Key: Filled circles -- OAO-2 ultraviolet fluxes \citep{79oao2}; open circles -- visual fluxes \citep{74dw};
  filled squares -- infrared fluxes from Table~\ref{tab:phot}; grey line -- Kurucz model atmosphere
  fluxes (Cohen, private communication).  The model fluxes are not a fit to the empirical fluxes.}
  \label{fig:flux}
\end{center}
\end{figure*}

\subsection{The Emergent Flux and Effective Temperature} \label{sec:fandt}

The emergent flux at the stellar surface ($\cal{F}$) and the effective
temperature ($T_{\mathrm{eff}}$) of a star are related by

\begin{equation}
\sigma T_{\mathrm{eff}} = {\cal{F}} =
\frac{4f}{\theta^{2}_{\mathrm{LD}}}  \label{eqn:fandteff}
\end{equation}
where $\sigma$ is the Stefan-Boltzmann radiation constant.

Substituting the bolometric flux and limb-darkened angular diameter determined
above into equation~(\ref{eqn:fandteff}) gives the emergent flux at the stellar surface
${\cal{F}} = (5.32\pm0.14)\times10^{8}$\,Wm$^{-2}$ and the effective temperature
$T_{\mathrm{eff}} = 9842\pm64$\,K.

The literature contains many estimates of the effective temperature of Sirius but the only
previous determination based on a measured angular diameter and empirically determined
flux distribution is that of \citet{76code}.  They obtained 9970$\pm$160\,K which lies
within one standard deviation from the result of significantly improved accuracy presented
here.

The Kurucz model atmosphere for Sirius which \citet{92cohen} used in their approach to
absolute stellar calibration had an effective temperature of 9850\,K in remarkable
agreement with the independent value established here.  Although use was made of this
model atmosphere in making allowance for absorption lines it is stressed that the current
work is essentially independent of theoretical models.  It therefore provides an independent
check on the effective temperature and angular diameter determined by \citet{92cohen}.  The
agreement for both quantities is remarkable with the present work providing results of
significantly improved accuracy.

\subsection{The Radius and Luminosity}

The angular diameter and bolometric flux can be combined with the parallax to determine the
stellar radius and the stellar luminosity respectively.  Combining the limb-darkened angular
diameter with the Hipparcos parallax \citep{07hip} of 379.21$\pm$1.58\,mas gives the radius
of Sirius A as $R/R_{\odot}$ = 1.713$\pm$0.009.

The luminosity is given by

\begin{equation}
L = 4\pi f\frac{C^{2}}{\pi^{2}_{\mathrm{p}}}
\end{equation}
where $C$ is the conversion from parsecs to metres.  Substitution
leads to $L/L_{\odot} = 24.7\pm0.7$.

The basic data and derived parameters for Sirius are listed in Table~\ref{tab:params}.

\begin{table}
\begin{center}
  \caption{The parameters of Sirius A. $\theta_{\mathrm{LD}}$ is the limb-darkened angular
  diameter, $\pi_{\mathrm{p}}$ is the parallax, $f$ is the bolometric flux received from the
  star, $\cal{F}$ the emergent flux at the stellar surface, $T_{\mathrm{eff}}$
  the effective temperature, $R/R_{\odot}$ the radius in solar radii and $L/L_{\odot}$
  is the luminosity in solar luminosities.}
  \label{tab:params}
  \vspace{2mm}
  \begin{tabular}{lccc}
  \hline
 \multicolumn{1}{c}{Parameter} & Value & $\pm$\% & Ref. \\
 \hline
 $\theta_{\mathrm{LD}}$ (mas) & 6.041$\pm$0.017 & 0.27 & 1 \\
 $\pi_{\mathrm{p}}$ (mas) & 379.21$\pm$1.58 & 0.42 & 2 \\
 $f$ (Wm$^{-2}$) & (1.141$\pm$0.029)$\times10^{-7}$ & 2.5 & 1 \\
 $\cal{F}$ (Wm$^{-2}$) & (5.32$\pm$0.14)$\times10^{8}$ & 2.6 & 3 \\
 $T_{\mathrm{eff}}$ (K) & 9842$\pm$64 & 0.65 & 3 \\
 $R/R_{\odot}$ & 1.713$\pm$0.009 & 0.5 & 3 \\
 $L/L_{\odot}$ & 24.7$\pm$0.7 & 2.8 & 3 \\
 \hline
 \end{tabular}
\end{center}
\medskip
References: (1) This work; (2) \citet{07hip}; (3) Derived.
\end{table}

\section{Summary}

A new accurate limb-darkened angular diameter for Sirius A has
been determined which is in excellent agreement with and confirms
the ESO/VLTI measurement by \citet{03ketal}.  The weighted mean of
the two determinations is $\theta_{\mathrm{LD}} = 6.041\pm0.017$\,mas
($\pm$0.27\,\%).  An accurate value for the bolometric flux has been
computed from flux-calibrated photometry and spectrophotometry.
Combination of the angular diameter with the bolometric flux and
the revised Hipparcos parallax has led to improved values for the
emergent flux at the stellar surface, the effective temperature, the
radius and the luminosity.  It is noted that the accuracies of the
emergent flux, effective temperature and luminosity are limited by
the accuracy of the calibration of the photometric fluxes.

The effective temperature and angular diameter are in excellent agreement
with those derived by \citet{92cohen} through spectral irradiance calibration
in the infrared.

The accurate value for the angular diameter of Sirius A is important for its role
as a primary infrared standard beyond 20\,$\mu$m as advocated by \citet{92cohen}.  In
 addition the improved values and accuracy of radius, effective temperature and luminosity
will provide tighter constraints on the evolutionary status of Sirius A than
has been possible to date.

\section*{Acknowledgements}

The SUSI programme is funded jointly by the Australian Research
Council and the University of Sydney.  Martin Cohen kindly provided
the flux distribution for the model atmosphere for Sirius that
Bob Kurucz had computed for him.  This work has made use of the SIMBAD database,
operated by CDS, Strasbourg, France and VizieR \citep{vizier}.





\end{document}